\documentclass[aps,showpacs]{revtex4}
\usepackage{amssymb}
\usepackage{graphicx}
\usepackage{dcolumn}
\usepackage{amsmath}

\begin{document}

\title{The Kondo effect in the presence of Van Hove singularities:
A numerical renormalization group study}
\author{A. K. Zhuravlev and V. Yu. Irkhin}
\email{Zhuravlev@imp.uran.ru, Valentin.Irkhin@imp.uran.ru}
\affiliation{Institute of Metal Physics, 620990 Ekaterinburg, Russia}

\begin{abstract}
A  numerical renormalization group (NRG) investigation
 of the one-centre $t-t'$  Kondo problem is performed for the square lattice
with account of
logarithmic Van Hove singularities (VHS) in the electron density of states.
The  magnetic susceptibility, entropy  and specific heat are calculated.
The temperature dependences of the
thermodynamic properties in the presence of VHS turn out to be non-trivial.
For finite $t'$   inverse logarithm of the corresponding Kondo temperature $T_K$
demonstrates a crossover from the square-root  to standard linear dependence
on the $s-d$ exchange coupling.
The  low-temperature behavior of magnetic susceptibility and  linear specific heat
are investigated, and the Wilson ratio is  obtained.
For $t'\rightarrow 0$ the Fermi-liquid behavior is broken.

\end{abstract}

\maketitle


\section{Introduction}

The Kondo effect is extensively studied starting from pioneering works
by Kondo \cite{552} who explained
the minimum of resistance in metallic alloys owing to
resonance $s-d$ scattering of conduction electrons by magnetic impurities.
The solution of the Kondo problem \cite{Wilson,Wiegmann,Hewson}
is a very beautiful chapter in the history of modern theoretical physics.

The Kondo phenomenon is a key for explaining the behavior of
heavy-fermion compounds and Kondo lattices \cite{Stewart,Brandt},
 non-Fermi-liquid (NFL) systems \cite{Maple,Stewart1},
anomalous electronic properties of
metallic glasses  \cite{Cox},  quantum dots \cite{qd}, and  other systems.
The Kondo anomalies are also studied in systems of reduced dimensionality where they
have a  number of experimental  peculiarities  \cite{Blachly}.
It is evident that the Kondo effect in such systems has a number of peculiar
features from the theoretical point of view too.

The Kondo effect owing to Cu$^{2+}$  spins in the CuO$_{2}$ planes
is observed in layered n-type  cuprates
(La,Ce)$_{2}$CuO$_{4}$, (Pr,Ce)$_{2}$Cu$_{4}$, and (Nd,Ce)$_{2}$CuO$_{4-\delta}$ \cite{Sekitani}.

Last time, the  Kondo effect in graphene (truly two-dimensional system with a peculiar electron spectrum)
is discussed \cite{Sengupta,Zhuang,Wehling}. Because of the pseudogap in the spectrum, the Kondo effect
for the undoped graphene
exists  under restricted conditions only, but  for a doped substance the Kondo phase
is present for all parameters \cite{Zhuang}.

We can mention also some layered $f$-systems where experimental investigations and first-principle band calculations demonstrate existence of two-dimensional features in electron properties. Here belong the compounds  CeCoIn$_5$ (where the layers CeIn are present) \cite{CeCoIn}, CeCoGe$_2$ \cite{CeCoGe}, CePt$_2$In$_7$ \cite{CePt}, CeRhIn$_5$, Ce$_2$RhIn$_8$ \cite{CeRhIn}, UCo$_{0.5}$Sb$_2$ (where two-dimensional weak localization is observed) \cite{UCo}.

The layered Kondo lattice
model was proposed for quantum critical beta-YbAlB$_{4}$ where
two-dimensional boron layers are Kondo coupled via interlayer Yb moments
\cite{YbAlB4}. CeRuPO seems to be one of the rare examples of a ferromagnetic
Kondo lattice where LSDA+U calculations evidence a quasi-two-dimensional
electronic band structure, reflecting a strong covalent bonding within the
CeO and RuP layers and a weak ioniclike bonding between the layers \cite
{CeRuPO}.

The above $f$-systems  demonstrate often both local-moment and itinerant-electron
features. Large linear specific heat and NFL behavior is observed
also in some $d$-systems including layered ruthenates Sr$_{2}$RuO$_{4}$ \cite{ruthenates} and Sr$_{3}$Ru$_{2}$O$_{7}$ \cite{ruthenates1}.
Besides correlation effects, anomalies of electron properties
in such systems are owing yo
the presence of Van Hove singularities near the Fermi level.

In the present work we treat the one-centre Kondo problem  with the singular electron density of states. Earlier this
problem was considered by Gogolin \cite{gogolin} who used a ``fast parquet'' approach. In fact, such complicated methods
are somewhat ambiguous, and the numerical renormalization group (NRG) results of paper \cite{zhuravlev} do not agree with
the results of Ref.\cite{gogolin}. Therefore we start in Sect. 2 from the standard perturbation theory and also apply
the \textquotedblleft poor man scaling\textquotedblright\ approach by Anderson \cite{And}.

In Sect. 3 we apply to the problem the NRG method, the technical details for our case being considered in Appendix. The
simple perturbation results for the Kondo temperature agree with NRG much better than the parquet results \cite{gogolin}.
The physical quantities in the presence of the logarithmic singularity near the Fermi level are investigated. We calculate
the magnetic susceptibility and specific heat, in particular at low temperatures, discuss the problem if universal behavior
and calculate the Wilson ratio.

\section{The Kondo model with Van Hove singularities}

We use the Hamiltonian of the one-centre $s-d(f)$ exchange (Kondo) model
\begin{equation}
H_{sd}=\sum_{\mathbf{k}\sigma }\varepsilon _{\mathbf{k}}c_{\mathbf{k}\sigma
}^{\dagger }c_{\mathbf{k}\sigma }^{{}} - \sum_{\mathbf{k}\mathbf{k^{\prime }}%
\alpha \beta } J_{\mathbf{k}\mathbf{k^\prime}}
\mathbf{S}\mbox {\boldmath
$\sigma $}_{\alpha \beta }c_{\mathbf{k}\alpha }^{\dagger }c_{\mathbf{%
k^{\prime }}\beta }^{{}} \label{Ham_Kondo}
\end{equation}%
where $\varepsilon _{\mathbf{k}}$ is the band energy, $\mathbf{S}$ are spin operators, $\sigma $ are the Pauli matrices,
in the case of contact coupling $J_{\mathbf{k}\mathbf{k^\prime}}=J/N_s$ where
$J$ is the $s-d(f)$ exchange parameter, $N_s$ is the number of lattice sites.

Kondo \cite{552} found that high-order perturbation contributions to various
physical properties contain logarithmically divergent corrections. As
demonstrated further investigations of the Kondo problem, there occurs
a pole at the boundary of strong coupling region, which is called the Kondo temperature
(in fact, this is a crossover scale). For a smooth density of states
$\rho (E)$ this quantity is estimated as
\begin{equation}
T_{K}\varpropto D\exp {}\frac{1}{2J\rho (0)},  \label{kons}
\end{equation}
where  $D$ is the half-bandwidth.
We treat the case of logarithmically divergent bare density of electron
states
\begin{equation*}
\rho (E)=A\ln \frac{D}{B|E+\Delta|}
\end{equation*}%
(the energy is referred to the Fermi level, the constants $A$, $B$ and $\Delta$ are determined by the band
spectrum).
The logarithmic divergence in $\rho (E)$ is typical for the two-dimensional
case (in particular, for the layered ruthenates). However, similar strong
Van Hove singularities can occur also in some three-dimensional systems like
Pd alloys and weak itinerant ferromagnets ZrZn$_{2}$ and TiBr$_{2}$ \cite%
{VKT,pickett}.

First we consider perturbation expansion for the resistivity, following to
the original approach by Kondo \cite{552}. We write down the inverse
transport relaxation time with the Kondo correction%
\begin{equation}
\tau ^{-1}(E)=\tau _{0}^{-1}(E)[1+4J\,g(E,0)].  \label{eq:6.37}
\end{equation}%
Here
\begin{equation}
g(E,T)=\sum_{\mathbf{k}}\frac{1/2-n_{\mathbf{k}}}{E-\varepsilon _{\mathbf{k}}%
}=\int dE^{\prime }\rho (E^{\prime })\frac{1/2-f(E^{\prime })}{E-E^{\prime }} \label{gET}
\end{equation}%
with $n_{\mathbf{k}}=f(\varepsilon _{\mathbf{k}})$ is the Fermi function.
After integration by part, in the case $\Delta =0$ we obtain to logarithmic
accuracy for the resistivity%
\begin{equation}
R_{sd}\sim \int dE\rho (E)\left( -\frac{\partial f(E)}{\partial E}\right) \,\tau _{0}^{-1}(E)\left[ 1-2AJ\int dE^{\prime
}\left( -\frac{\partial f(E^{\prime })}{\partial E^{\prime }}\right) \ln ^{2}\left|\frac{D}{E^\prime}\right|\right]
\end{equation}%
so that%
\begin{equation}
R_{sd}=R_{sd}^{(0)}\left( 1-2AJ\ln ^{2}\frac{D}{T}\right) ,\qquad R_{sd}^{(0)}\sim J^{2}S(S+1)A\ln \frac{D}{T}
\label{eq:6.6}
\end{equation}%
%
%
Applying the Abrikosov--Suhl summation (see \cite{552}) we get
\begin{equation}
R_{sd}=R_{sd}^{(0)}\left( 1+JA\ln ^{2}\frac{D}{T}\right) ^{-2} \label{eq:6.7}
\end{equation}%
which yields a non-standard expression for the Kondo temperature,
\begin{equation}
T_{K}\simeq D\exp\left[-\left|\frac{1}{AJ}\right|^{1/2}\right] . \label{perturb1}
\end{equation}

\begin{figure}[htbp]
\includegraphics[width=3.3in, angle=0]{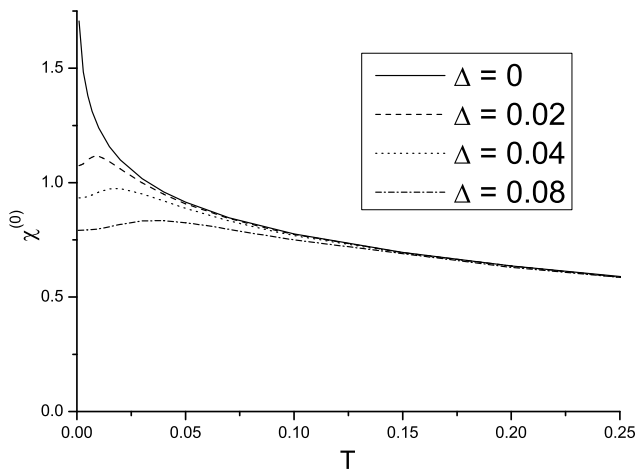}
\caption{The temperature dependence of non-interacting magnetic susceptibility for conduction electrons $\chi^{(0)}(T)$ at
$\Delta=0$, 0.02, 0.04, 0.08} \label{Fig_chi0}
\end{figure}

We calculate also the Kondo corrections to the static impurity magnetic susceptibility by generalizing
consideration of Ref.\cite{552} to the case of singular density of states.
Expanding to second order in $J$ we derive (cf. \cite{552,IKFTT,Hewson})
\begin{equation}
\chi (T)=\,\frac{S(S+1)}{3T}\left[ 1+2J\chi ^{(0)}-2J^{2}\sum_{\mathbf{kk}%
^{\prime }}\frac{n_{\mathbf{k}}(1-n_{\mathbf{k}^{\prime }})}{(\varepsilon _{%
\mathbf{k}}-\varepsilon _{\mathbf{k}^{\prime }})^{2}}\right]   \label{Sef}
\end{equation}%
where%
\begin{equation}
\chi ^{(0)}(T)=-\sum_{\mathbf{k}}\frac{\partial n_{\mathbf{k}}}{\partial
\varepsilon _{\mathbf{k}}}=\int dE\rho (E)\left( -\frac{\partial f(E)}{%
\partial E}\right) \simeq A\ln \frac{D}{\max (\Delta -T,T)}  \label{hi0}
\end{equation}%
is the Pauli susceptibility of non-interacting conduction electrons with the
singular density of states, which is shown in Fig.\ref{Fig_chi0}.
For $\Delta \neq 0$ this quantity has a maximum
at $T\simeq \Delta/2 $. Such a behavior is typical for the case where a
density-of-states peak is present near the Fermi level \cite{II}.

Performing integration and summation of the series of logarithmic terms we
have%
\begin{equation}
\chi (T)=\,\frac{S(S+1)}{3T}\left[ 1+\frac{2J\chi ^{(0)}}{1+J\chi ^{(0)}\ln (D/T)}\right]   \label{hiparrq}
\end{equation}%
which yields at $\Delta =0$ the same result for the Kondo temperature (\ref%
{perturb1})

The correction to magnetic impurity entropy can be written down in an analogous way
to obtain  \cite{552,Hewson}%
\begin{equation}
\mathcal{S}_\mathrm{imp}(T)=\ln (2S+1)+\frac{\pi ^{2}}{3}\frac{S(S+1)\left[ 2J\chi ^{(0)}%
\right] ^{3}}{[1+J\chi ^{(0)}\ln (D/T)]^{3}}  \label{entr}
\end{equation}%
Then we have for impurity specific heat

\begin{equation}
C_\mathrm{imp}(T)=T\frac{d\mathcal{S}_\mathrm{imp}(T)}{dT}=\frac{\pi ^{2}S(S+1)}{[1+J\chi ^{(0)}\ln
(D/T)]^{4}}\left( \left[ 2J\chi ^{(0)}\right] ^{4}+8J^{3}T\frac{d\chi ^{(0)}%
}{dT}\left[ \chi ^{(0)}\right] ^{2}\right)   \label{cimp}
\end{equation}%
The first term in the brackets yields the same structure as in the case of a smooth density of states (where $\chi
^{(0)}=\rho $ and the singular contribution occurs in the fifth order in $J$ only), and the second term is owing to logarithmic singularity in $\chi ^{(0)}$.
Being of the third order in $J$, the latter term can dominate. Beside that, it can change its sign and become negative; as we shall see below in Sect.3, this is important for the $C_\mathrm{imp}(T)$ behavior.

Of course, the above expressions are applicable for $T>T_{K}$ only.

To perform a more formal consideration, we can apply the \textquotedblleft poor man scaling\textquotedblright\ approach
\cite{And}. This treats the dependence of effective (renormalized) model parameters on the cutoff parameter $C$ which
occurs at picking out the singular contributions from the Kondo corrections to the effective coupling $J_{ef}(C)$ (with
$J_{ef}(-D)=J$). The second-order singular correction to $J_{ef}(C)$ can be obtained in the form \cite{kondo}
\begin{equation}
\delta J_{ef}=-2J^{2}\sum_{\mathbf{k}}\frac{n_{\mathbf{k}}}{E-\varepsilon _{%
\mathbf{k}}}
\end{equation}%
Picking out in the sum the contribution of intermediate electron states near the Fermi level with
$C<\varepsilon_{\mathbf{k}}<C+\delta C$
we obtain
\begin{equation}
\delta J_{ef}(C)=2AJ^{2}\ln \frac{D}{|C+\Delta|}\frac{\delta C}{C}  \label{ief}
\end{equation}
Replacing in the right-hand part  $J\rightarrow $ $J_{ef}(C)$ we obtain%
\begin{equation}
\frac{\partial }{\partial C}\frac{1}{J_{ef}(C)}=-\frac{2A}{C}\ln \frac{D}{|C+\Delta|} \label{oneI}
\end{equation}
Solving this equation for $T_K \gg\Delta$ we get
\begin{equation}
\frac{1}{J_{ef}(C)}=\frac{1}{J}+A\ln^2\left|\frac{D}{C}\right|. \label{1/I}
\end{equation}
Then we obtain from the condition $1/J_{ef}(T_{K})=0$ again the result (\ref{perturb1}).

One can see that the expression (\ref{perturb1}) is different from the
corresponding parquet result \cite{gogolin}
\begin{equation}
T_{K}\simeq D\exp \left[-\left|\frac{2}{AJ}\right|^{1/2}\right].   \label{gogolin}
\end{equation}

However,  the NRG
calculations \cite{zhuravlev} confirms the perturbation expression
(\ref{perturb1})  rather than (\ref{gogolin}) (see the discussion below). The corresponding problems of the parquet
approximation in the Hubbard model are discussed in the works \cite{parquet}.

\section{Results of numerical calculations }

We consider $S=1/2$ Kondo model for the square lattice with the spectrum
\[
\varepsilon _{\mathbf{k}}=2t(\cos k_{x}+\cos k_{y})+4t^{\prime }(\cos
k_{x}\cos k_{y}+1)
\]%
The corresponding density of states is%
\begin{equation}
\rho (E)=\frac{1}{2\pi ^{2}\sqrt{t^{2}+Et^{\prime }-4t^{\prime 2}}}K\left(
\sqrt{\frac{t^{2}-E^{2}/16}{t^{2}+Et^{\prime }+4t^{\prime 2}}}\right) \simeq
\frac{1}{2\pi ^{2}\sqrt{t^{2}-4t^{\prime 2}}}\ln \frac{16\sqrt{%
t^{2}-4t^{\prime 2}}}{|E+8t^{\prime }|}
\end{equation}%
where $K(E)$ is the complete elliptic integral of the first kind. In the numerical calculations, the Fermi level is
supposed to be located in the band centre, so that the band is determined by $|E|<D=4|t|$. The distance from the Van Hove
singularity to the Fermi level is $\Delta=8t^{\prime }$. When presenting numerical results, we put $D=1$.

For $t^{\prime }=0$ we derive%
\begin{equation}
\rho (E)=\frac{2}{\pi ^{2}D}K\left( \sqrt{1-\frac{E^{2}}{D^{2}}}\right)
\simeq \frac{2}{\pi ^{2}D}\ln \frac{4D}{|E|},
\end{equation}%
so that, according to (\ref{perturb1}),
\begin{equation}
T_{K}\simeq D\exp \left[ -\left| \frac{\pi ^{2}D}{2J}\right| ^{1/2}%
\right] .  \label{perturb}
\end{equation}

The details of 
NRG calculations are discussed in Appendix.

The Kondo temperature in NRG calculations is determined from the temperature dependence of impurity
magnetic susceptibility
$\chi_{\mathrm{imp}}(T)$ by using condition $T_K\chi_{\mathrm{imp}}(T_K) = 0.0701$ that is standard in the NRG method \cite{Wilson}.

\begin{figure}[htbp]
\includegraphics[width=3.3in, angle=0]{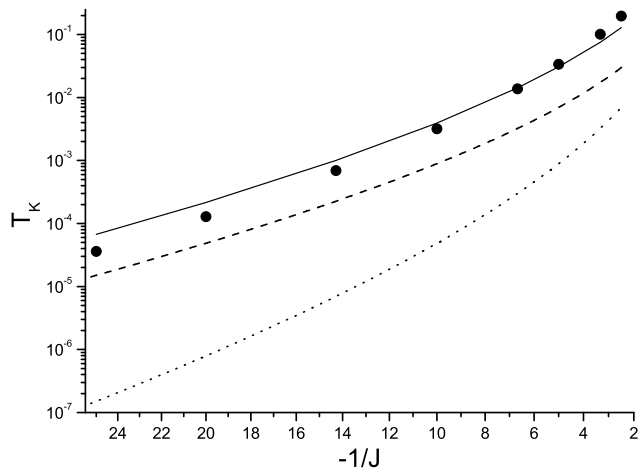}
\caption{The dependence $T_K(J)$ for $t'=0$. Circles are NRG results,  solid line corresponds to the Nagaoka-Suhl formula
(\ref{eq:NS}), dashed line to Eq.(\ref{perturb}) and dotted one to Eq.(\ref{gogolin}) (without any fitted preexponential
constants)} \label{Fig_TKt0}
\end{figure}

\begin{figure}[htbp]
\includegraphics[width=4.3in, angle=0]{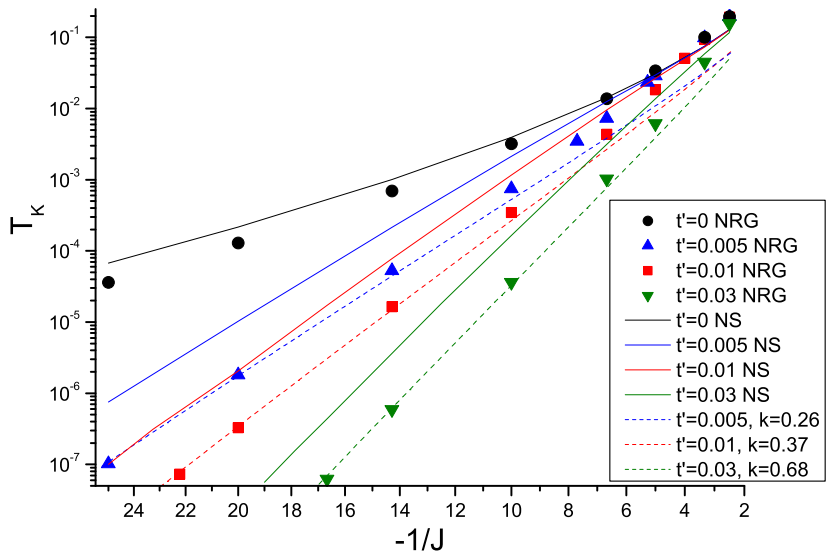}
\caption{Fitting of the dependences  $T_K(J)$ for  $t'=0,0.005, 0.01, 0.03$. Circles, top-up triangles,  squares and top-down triangles are corresponding NRG results, solid
lines (from above to below)  correspond to the Nagaoka-Suhl formula (\ref{eq:NS}), dashed lines (from above to below) show the two-loop  result (\ref{two-loop}) with fitted constants $k$ } \label{Fig_TK}
\end{figure}

Figs.\ref{Fig_TKt0}, \ref{Fig_TK}  show the  dependence of the Kondo temperature on the bare coupling  $|J|$ in the logarithmic scale. One can see that for
finite $t'$ there occurs a crossover with decreasing $|J|$ from the square-root dependence (\ref{perturb1}) to the standard
Kondo behavior (\ref{kons}). This crossover is qualitatively described by the lowest-order scaling.

It  should  be  noted  that the Kondo phenomenon should be invariant under particle-hole transformation ($\varepsilon
_{\mathbf{k}} \rightarrow -\varepsilon _{\mathbf{k}}$, i.e., $\rho(\epsilon) \rightarrow \rho(-\epsilon)$). However, in the
case of a non-symmetric density of states  using the formula $1=2Jg(T_K,0)$ (where $g(E,T)$ is determined by (\ref{gET}))
yields slightly  different results for the peak above and below the Fermi level. Therefore we use the Nagaoka-Suhl formula
(see \cite{552})
\begin{equation}
1=2Jg(0,T_K) \label{eq:NS}
\end{equation}
which works somewhat better at large $J$, but  slightly worse at small  $J$.
(In fact, all such approximations yield nearly the same result at small $J$.)

According to (\ref{oneI}), for $\Delta \ll T_K$ we have $\ln T_K \sim 1/\sqrt{|J|}$ and for $\Delta \gg T_K$ we come to the
standard linear behavior $\ln T_K \sim 1/|J|$ with enhanced $\rho(0)$. For intermediate  $J$, a crossover takes place.

The lowest-order scaling describes satisfactorily the numerical data at large $|J|$. However, it is insufficient to fit
the numerical results at small $|J|$.
Therefore we use the two-loop scaling result \cite{Wilson,Hewson}
\begin{equation}
T_{K}=kD|2J\rho |^{1/2}\exp \left[ 1/2J\rho \right] \label{two-loop}
\end{equation}
This expression implies that only small vicinity of the Fermi surface with nearly constant $\rho(E)\simeq\rho$ works.
This assumption becomes not valid with increasing $|J|$ since the whole logarithmic peak becomes important.
It should be noted that, as follows from the structure of scaling equations (see Ref.\cite{Hewson}),
the factor of $|2J\rho|^{1/2}$ does not occur in the expression for $T_K$ at $\Delta=0$.

The fitted constant $k$ should be determined by the whole form of the function $\rho (E)$.
At small $t'$, $k$ is small since the quantity $\rho (0)$ is large owing to peaks.
On the contrary, for larger $t'=0.03-0.05$, $\rho (0)$ becomes small.

Since $T_K$ is high in our case of singular density of states, consideration of the situation with $\Delta  \sim T_K$ is
quite realistic. We can see that appreciable deviations owing to the singularity occur even if $\Delta $ is not too small
and exceeds $T_K$.

The magnetic susceptibility owing to impurity can be expressed as a difference of magnetic susceptibilities of the whole
system and the system without impurity:
\begin{equation}
\label{chi_imp}\chi(T)\equiv \chi_\mathrm{imp}(T)=\chi_\mathrm{tot}(T)-\chi_\mathrm{band}(T) ,
\end{equation}
where $\chi_\mathrm{tot}$ is the total magnetic susceptibility, and $\chi_\mathrm{band}=2N_s \chi^{(0)}$ is the susceptibility of
non-interacting band electrons (for two spin projections).
Apart from the susceptibility (\ref{chi_imp}), the so-called local magnetic susceptibility
$\chi_\mathrm{loc}$ is frequently introduced as well:
\begin{equation}
  \label{ChiLocal}
    \chi_\mathrm{loc}(T) = \int\limits_0^{1/T}\langle S_z(\tau)S_z\rangle d\tau \ ,
\end{equation}
This is the susceptibility of a \emph{single} impurity in a magnetic field that
acts locally, only on this impurity; its magnitude, therefore, can hardly be measured experimentally, in contrast to $\chi_\mathrm{imp}$.

In principle, $\chi_{\mathrm{imp}}$ and $\chi_{\mathrm{loc}}$ can behave quite differently. This possibility was mentioned
in Ref.\cite{Santoro}, where the reason for the difference is related to the energy dependence $\rho(E)$ and is asserted
that this disappears for a flat band of half-width $D$ in the limit of $D\rightarrow\infty$.

The impurity entropy and specific heat are defined in a similar way.

\begin{figure}[htbp]
\includegraphics[width=3.3in, angle=0]{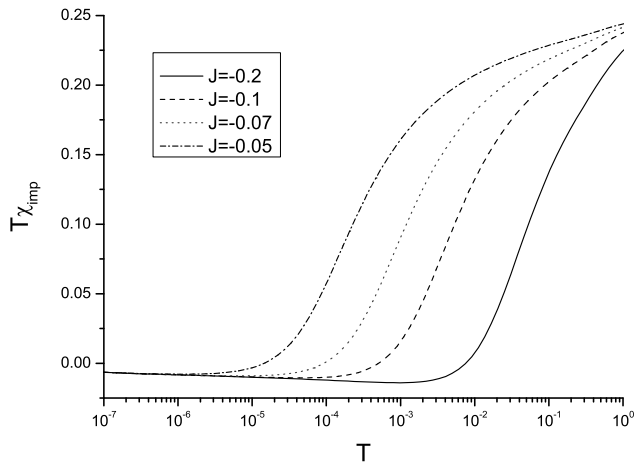}
\caption{The temperature dependence $T\chi_\mathrm{imp}(T)$ for $t'=0$ and $J=-0.2, -0.1, -0.07, -0.05$ (lines from below
to above)} \label{Fig_Thi_t0}
\end{figure}

\begin{figure}[htbp]
\includegraphics[width=3.3in, angle=0]{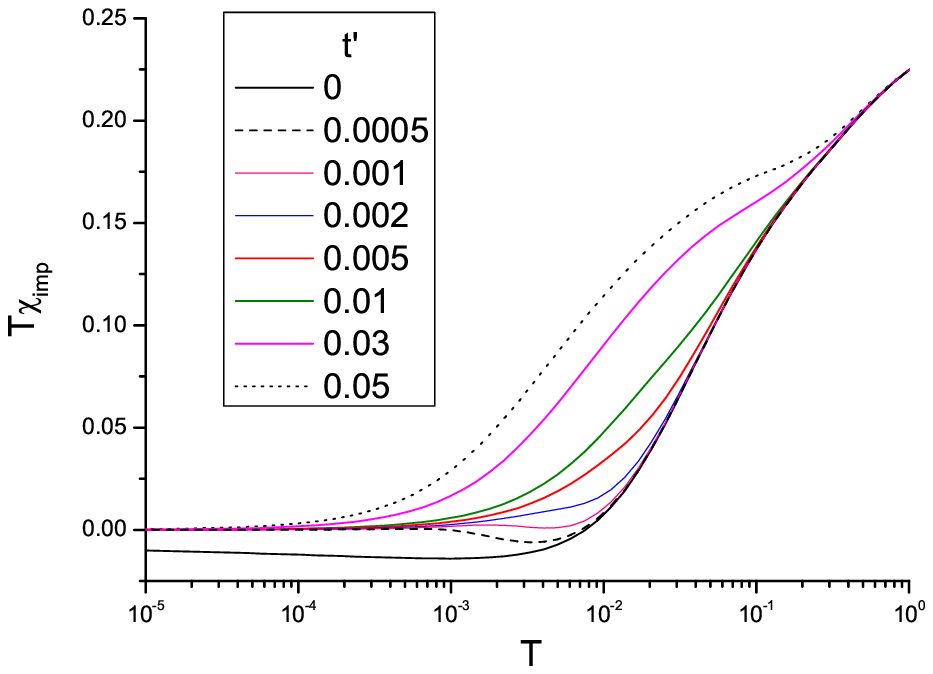}
\caption{The temperature dependence $T\chi_\mathrm{imp}(T)$ for $J=-0.2$ and  different $t'=0$, 0.0005, 0.001, 0.002,
0.005, 0.01, 0.03, 0.05 (lines from below to above)} \label{Fig_ThiJ02}
\end{figure}

\begin{figure}[htbp]
\includegraphics[width=3.3in, angle=0]{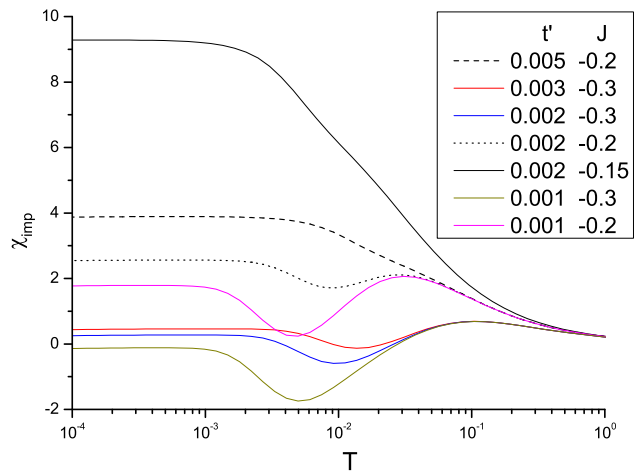}
\caption{The temperature dependence $\chi_\mathrm{imp}(T)$ for different $J$ and  $t'$}
\label{Fig_hiJ02}
\end{figure}

Figs.\ref{Fig_Thi_t0}-\ref{Fig_hiJ02} show the temperature dependence of magnetic susceptibility 
for different $t'$ values. At
high temperatures, $\chi (T)$ obeys the Curie law, and at low temperatures it demonstrates the Pauli behavior (except for the case $\Delta\rightarrow 0$).

For the standard flat-band case we have  $T\chi(T)= \phi(T/T_K)$, so that the curves $T\chi(T)$ are universal: a change in
$J$ results in a change of $T_K$ only. In our situation, such a simple universality does not hold. In particular,  for
$t'=0$  this fact (illustrated by Fig.\ref{Fig_Thi_t0}) was demonstrated in Ref.\cite{zhuravlev}.


One can see from Fig.\ref{Fig_hiJ02} that the minimum of $\chi (T)$ occurs with decreasing $\Delta$, its position being
determined by the maximum of $\chi^{(0)}(T)$ in Eq.(\ref{hi0}).
This minimum is due to the strong  energy dependence of the bare density of states, see
Eqs.(\ref{Sef})-(\ref{hiparrq}). Thus, besides the Kondo temperature $T_K$, a second characteristic scale can occur in our
problem, which is determined by $\Delta$. Therefore we have to discuss the meaning of the Kondo temperature $T_K$ in more
detail. This quantity comes from the expansion in $J$ starting from high temperatures where impurity susceptibility $\chi
(T)$ obeys the Curie law. $T_K$ is determined as a temperature scale where a considerable deviation from this law occurs.
In the  flat-band situation and for smooth $\rho(E)$, $\chi (T)$ decreases monotonously with increasing $T$, the universal
behavior taking place. This picture is characterized by the  ratio $w=4T_K\chi(0)\approx0.41$ \cite{Wilson,Wiegmann,Hewson}
which relates high- and low-temperature scales $T_K$ and $\chi (0)$. Such a behavior holds for $T_K<\Delta$, but for small
$\Delta$ values the universality is broken, so that the intermediate-temperature dependence $\chi (T)$ becomes complicated
and $w$ deviates from 0.41 (see Table \ref{Tabl_ChiTK}).

The Wilson ratio $R = (4\pi^2/3)\chi_\mathrm{imp}(0)/ \gamma_\mathrm{imp}$
relating low-temperature susceptibility and linear specific heat $\gamma_\mathrm{imp}$
is also  presented in the Table \ref{Tabl_ChiTK}.
It is important  that even  for rather small $\Delta$ the value of $R$ is still close to 2.
Therefore a Fermi-liquid behavior, characteristic for the Kondo problem \cite{Wilson},
is restored at low temperatures, except for the case of extremely
small $\Delta$ where $\chi_\mathrm{imp} (0)$ and $C_\mathrm{imp}(0)$ can even become negative.

\begin{table}[htb]
\caption{Results of our NRG calculations for different $t'$ and $J$:
$T_K$ (first line); Wilson ratio $R$ (second line); the quantity $w = 4 T_K\chi(0)$ (third line; Wilson's value is $w = 0.4128\pm0.002$)}
\begin{tabular}{|c|c|c|c|c|}
\hline
  $t'$ &  $J=-0.1$ & $J=-0.15$ & $J=-0.2$ &  $J=-0.3$  \\  \hline
 &   1.19$\cdot10^{-5}$ & 4.76$\cdot10^{-4}$ & 0.00333 &  0.0277    \\
0.05 &   2.008 & 2.008 &2.014 &  1.99            \\
 &  0.414 & 0.416 & 0.416 & 0.415   \\ \hline
 & 3.62$\cdot10^{-5}$ & 0.00103 & 0.00613 & 0.0450  \\
0.03 & 1.995 & 1.988 & 2.005 & 1.98       \\
 & 0.413 & 0.413 & 0.414 & 0.413  \\ \hline
 &   3.45$\cdot10^{-4}$ & 0.00436 & 0.0184 & 0.0930  \\
0.01 &  1.998 & 1.995 &1.991 &    2.12            \\
 &   0.415  & 0.416 & 0.425 & 0.386   \\ \hline
 &  7.49$\cdot10^{-4}$ & 0.00726 & 0.0288 & 0.0989   \\
0.005 &  1.999 & 1.997 &2.069 &   2.237     \\
 & 0.416 & 0.418 & 0.447 & 0.268736  \\ \hline
 & 0.00112 & 0.0102 & 0.0320 &   0.100007        \\
0.003 & 2.002 & 1.992 & 1.998 & 3.18      \\
 & 0.417 & 0.449 & 0.392   & 0.184   \\ \hline
 &  0.00144 & 0.0121  &  0.0330  & 0.1004       \\
0.002 &  1.996 & 1.97  &  2.040  & -177$^a$      \\
 & 0.417 & 0.44768 & 0.337 & 0.108            \\ \hline
 & 0.00208 & 0.0134 & 0.0337 & 0.1006       \\
0.001 & 1.989 & 2.03 & 2.41 &  0.343$^b$    \\
 & 0.43 & 0.38 & 0.24 & -0.048            \\ \hline
 \end{tabular}

$^a$ $\chi(0) =0.27$,  $\gamma_\mathrm{imp}=-0.02$

$^b$ $\chi(0) =-0.12$,  $\gamma_\mathrm{imp}=-4.6$
\label{Tabl_ChiTK}
\end{table}

Occurrence of negative values of $\chi_\mathrm{imp}$ and ${\mathcal S}_\mathrm{imp}$ was demonstrated in
Ref.\cite{zhuravlev} by a strict analytical consideration of the simple  case $J=-\infty$ for the semielliptic density of
states. This is a common property of systems with  very narrow  density of states peaks near the Fermi level. Of course,
the quantities $\chi_\mathrm{tot}$ and $\chi_\mathrm{band}$ remain positive.

In the standard situation of Kondo effect one has $T\chi(T) = O(T)$, so that the impurity moment (note that $T\chi =
\langle S_z^2\rangle_\mathrm{tot} - \langle S_z^2\rangle_\mathrm{band}$, see Appendix) is completely compensated by
conduction electrons. At the same time, in the pseudogap situation (low density of states near $E_F$, $\rho(E) \propto
|E-E_F|^r, r>0$) the screening is incomplete, $T\chi(T)>0$ \cite{Ingersent}. Our problem describes the opposite situation:
$\rho(E\rightarrow 0)$ diverges and we have overcompensation: $T\chi(T)<0$.

\begin{figure}[htbp]
\includegraphics[width=3.3in, angle=0]{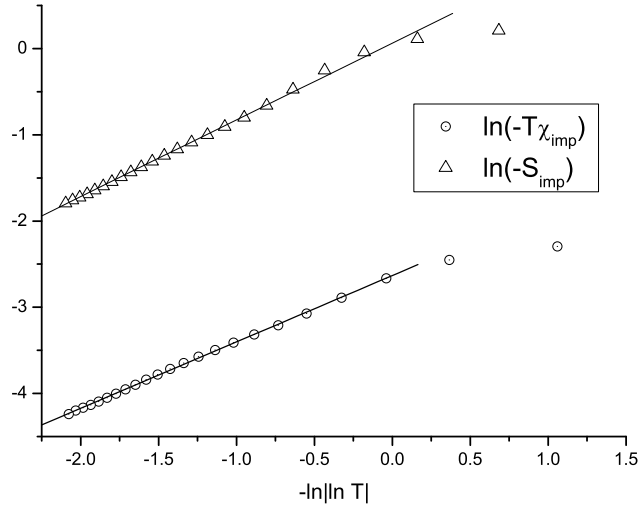}
\caption{The fitting of thermodynamic properties at low temperatures for  $\Delta=0$} \label{zhur_fig_10}
\end{figure}

The singular case $\Delta\rightarrow 0$ demonstrates essentially non-Fermi-liquid behavior (divergence of impurity magnetic
susceptibility  and   specific heat at  $T \rightarrow 0$). This situation of singular density of states at the Fermi level
is somewhat similar to the overscreened Kondo problem \cite{Col} where the number of scattering channel of conduction
electrons $n>2S$ ($S$ is the localized  spin value); in this case $\chi(T)$ demonstrates power-law behavior, but remains
positive.
The low-temperature behavior at  $\Delta=0$  can be fit  as \cite{zhuravlev}
\begin{eqnarray}
 \chi_\mathrm{imp} = -\frac{A}{T|\ln (T/D)|^\alpha} \ , A\approx0.072, \alpha\approx0.77
 \label{formula} \\
 \mathcal{S}_\mathrm{imp} = -\frac{B}{|\ln (T/D)|^\delta} \ , B\approx1.065, \delta\approx0.89.
 \label{formulaS}
\end{eqnarray}
(see Fig.\ref{zhur_fig_10}), so that
\begin{equation}
 C_\mathrm{imp} = -\frac{B\delta}{|\ln (T/D)|^{\delta+1}} .
 \label{formulaC}
\end{equation}
These asymptotics  are independent of $J$ (cf. Fig.\ref{Fig_Thi_t0}) since  at sufficiently low temperatures any value of $|J|$ manyfold exceeds both the temperature and the width of the infinitely thin logarithmic peak in $\rho(E)$.

According to Ref.\cite{under},  the NFL behavior with  $C \propto 1/\ln ^{4}(T_K/T)$ takes place 
in the case of
underscreened ($S>1/2$) Kondo problem  \cite{under}. 

\begin{figure}[htbp]
\includegraphics[width=3.3in, angle=0]{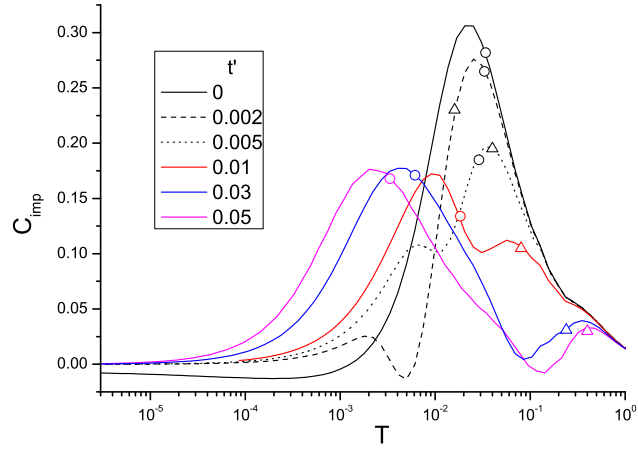}
\caption{The temperature dependence of impurity specific heat for $J=-0.2$ and different $t'=0$, 0.002, 0.005, 0.01, 0.03,
0.05 (lines from below to above, if one sees the left-hand part of the figure); circles mark $T_K$, and triangles $\Delta$}
\label{Fig_C_J02}
\end{figure}

Fig. \ref{Fig_C_J02} shows the temperature dependence of impurity specific heat
$C_\mathrm{imp}=C_\mathrm{tot}-C_\mathrm{band}$ for different $t'$. As a rule, this dependence demonstrates two peaks. At
not too small $\Delta$, the high-temperature maximum occurs at the temperature, determined by the distance from VHS to
$E_F$. This is owing to the non-monotonous dependence of $\chi^{(0)} (T)$, see Eqs.(\ref{hi0}), (\ref{cimp}). When
decreasing temperature and passing this maximum, $C_{\rm imp}(T)$ acquires a minimum and even can become negative. The
low-temperature peak is owing to the Kondo effect and takes place in the standard flat-band situation too (see Ref.
\cite{Oliveira}). One can see that its position corresponds roughly to the Kondo temperature. For  small $\Delta<T_K$, the
order of positions of the maxima becomes interchanged.

\begin{figure}[htbp]
\includegraphics[width=3.3in, angle=0]{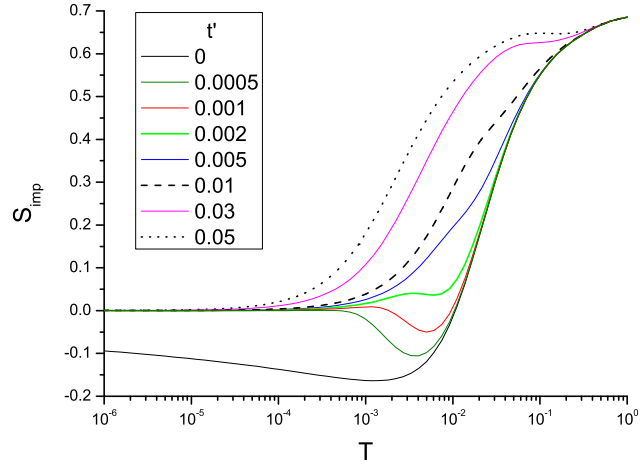}
\caption{The temperature dependence of magnetic entropy for $J=-0.2$ and different $t'=0$, 0.0005, 0.001, 0.002, 0.005,
0.01, 0.03, 0.05 (lines from below to above)} \label{Fig_S_J02}
\end{figure}

The corresponding magnetic entropy $\mathcal{S}_\mathrm{imp}$ is shown in Fig. \ref{Fig_S_J02}. One can see that this quantity
tends to the value $\ln 2=\ln (2S+1)$ at high temperatures and demonstrates the Kondo compensation at
low temperatures. The behavior turns out to be non-monotonous due to the maximum in $\chi^{(0)} (T)$ (see Eq.(\ref{entr})).

\section{Conclusions}

In the general problem of  magnetism of metals and alloys, the peaks in the bare density
of states (which are usually owing to Van Hove singularities) near the Fermi level play a
crucial role.
The  correlation  effects can result in a considerable qualitative and
quantitative     modification   of   the   corresponding   temperature
dependences \cite{VKT}. They can be important, e.g., for transition-metal alloys.

The considered $t-t'$  Kondo problem which includes VHS is a non-trivial example of influence
of density-of-states peaks on electron properties being combined with correlation effects \cite{II,VKT}.
Our treatment gives an example of exact numerical  solution of such a problem.
The resulting temperature dependences of thermodynamic properties include both one-particle
effects, connected with VHS, and many-electron Kondo features. Therefore the overall picture is rather complicated. At low temperatures, the Fermi-liquid behavior is restored, except for the case of
very small $t'$. In the latter case, a non-Fermi-liquid behavior takes place which should be studied by more advanced methods.

It would be also of interest to perform similar calculations
for the Kondo lattice problem, e.g., in some ``mean-field'' approximation.
A ``poor man scaling'' approach was applied to this problem in Ref.\cite{kondovh}.

The research described was supported in part by the Program
\textquotedblleft Quantum Physics of Condensed Matter\textquotedblright\
from Presidium of Russian Academy of Sciences.
The author are grateful to A.O. Anokhin and A.A. Katanin for useful discussions.

\section*{Appendix. Numerical renormalization group approach for the singular density of states}

Here we discuss some important details
of the numerical renormalization group (NRG)  method \cite{Wilson,Pruschke}
as applied to our problem of the singular density of states.

\subsection*{Construction of the Wilson chain}

\begin{figure}[ht]
\begin{center}
\unitlength 0.8cm
\begin{picture}(14,2)

\put(1,1){\circle*{0.2}}
\put(3,1){\circle{0.2}}
\put(5,1){\circle{0.2}}
\put(7,1){\circle{0.2}}
\put(9,1){\circle{0.2}}

\put(2,1){\vector(1,0){0.8}}
\put(2,1){\vector(-1,0){0.8}}
\put(4,1){\vector(1,0){0.8}}
\put(4,1){\vector(-1,0){0.8}}
\put(6,1){\vector(1,0){0.8}}
\put(6,1){\vector(-1,0){0.8}}
\put(8,1){\vector(1,0){0.8}}
\put(8,1){\vector(-1,0){0.8}}

\put(0.8,0.5){imp}
\put(1.8,1.3){$J$}
\put(2.8,0.5){$\epsilon_0$}
\put(3.8,1.3){$\gamma_{0}$}
\put(4.8,0.5){$\epsilon_1$}
\put(5.8,1.3){$\gamma_{1}$}
\put(6.8,0.5){$\epsilon_2$}
\put(7.8,1.3){$\gamma_{2}$}
\put(8.8,0.5){$\epsilon_3$}

\put(10,1.0){$\dots$}
\end{picture}
\end{center}
\caption{Representation of the Kondo model in the form of a
semiinfinite Wilson chain} \label{Fig_chain}
\end{figure}
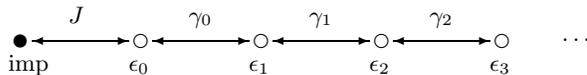

Following to Wilson \cite{Wilson} we use a unitary transformation to pass from the operators $c_{\mathbf{k}}$ to the operators $f_n$.
Then the impurity model with a Hamiltonian of type (\ref{Ham_Kondo}) is reduced to a semiinfinite chain (Fig.\ref{Fig_chain}) with a Hamiltonian of type:
\begin{eqnarray}
H_{sd}&=& -J\left[S^+f_{0\downarrow}^{\dagger }f_{0\uparrow} + S^-f_{0\uparrow}^\dagger f_{0\downarrow}
 + S_z\left(f_{0\uparrow}^\dagger f_{0\uparrow} - f_{0\downarrow}^\dagger f_{0\downarrow}\right)\right]   \nonumber \\
&+& \sum_{\sigma, n=0}^\infty \left[
               \epsilon_n f_{n\sigma}^\dagger f_{n\sigma}
               + \gamma_n \left( f_{n\sigma}^\dagger f_{n+1\sigma}
                  + f_{n+1\sigma}^\dagger f_{n\sigma}\right)\right] \ ,
\label{eq:H_chain}
\end{eqnarray}

The renormalization group procedure starts from the solution of the isolated-impurity problem (sites ``imp'' and
$\epsilon_0$ in Fig.~\ref{Fig_chain}).
At the first step, we add a first conducting electronic site $\epsilon_1$, and construct and diagonalize a Hamiltonian
matrix on this Hilbert space (with a fourfold higher dimensionality). This procedure is multiply repeated. However, since
the dimensionality of Hilbert space grows as $4^N$ ($N$ is the order number of an iteration), it is impossible to store all
the eigenstates during the calculation. Therefore, it is necessary to retain after each iteration only the states with the
lowest energies. If we restrict ourselves to a certain maximum number of stored states (determined by the computational
possibilities), it is necessary, starting from a certain iteration, to leave of the order of 1/4 of states at each step.

Unfortunately,  direct application of this scheme fails, since the disturbance introduced by the elimination of the
high-lying states proves to be too large. Wilson found a method of overcoming this difficulty. This reduces to the
artificial introduction of an exponential suppression of matrix elements $\gamma_n$, which decreases the coupling between
the retained and eliminated states, thereby decreasing the influence of the eliminated states. To this end, Wilson
\cite{Wilson} used a logarithmic discretization of the conduction band, i.e., replacement in (\ref{Ham_Kondo}) of an energy
range $\varepsilon _{\mathbf{k}} \in [\eta D\Lambda^{-m} , \eta D\Lambda ^{-m+1}]$, $\eta=1,-1$, $m=1,2,3,...$ by a single
level with an energy $\bar{\varepsilon} _{\eta m}$ equal to the average energy of this interval ($D$ is the half-width of
the conduction band, $\Lambda>1$). This results in a change in the density of states:
\begin{equation}
\rho(\varepsilon) \rightarrow \sum_{\eta,m} \alpha_{\eta m}^2 \delta(\varepsilon - \bar{\varepsilon}_{\eta m}) \ ,
\label{rho_discr}
\end{equation}%
where $$\alpha_{\eta m}^2 = \eta\int_{\eta D\Lambda^{-m}}^{\eta D\Lambda^{-m+1}}\rho(\varepsilon)d\varepsilon \ .$$ As a
result, the jumps will have the required decay, $\gamma_n\propto \Lambda ^{-n/2}$; for a flat band, Wilson obtained
analytically
\begin{equation}
\gamma_n =  \frac{D\left( 1+ \Lambda^{-1} \right) \left(1-\Lambda^{-n-1}\right) }{2\sqrt{1-\Lambda^{-2n-1}}
\sqrt{1-\Lambda^{-2n-3}}}\, \Lambda^{-n/2} \ , \epsilon_n=0 \ .
\label{gamma}
\end{equation}
In more complicated situations, the construction of the Wilson chain must be performed numerically. Usually (see, e.g.,
Ref.\cite{Chen}) this is performed in spirit of the initial work \cite{Wilson}, by numerical reproducing  Wilson's
analytical flat-band procedure for an arbitrary density of states. However, there exists another way which yields
equivalent results, but seems to be more natural. To bring the Hamiltonian matrix
\begin{eqnarray}
H_{\sigma} = \sum_{\eta,m} \bar{\varepsilon}_{\eta m} c_{\eta m,\sigma }^{\dagger }c_{\eta m,\sigma } \label{H_c}
\end{eqnarray}
to the tridiagonal form, one can use the Lanczos tridiagonalization algorithm \cite{Parlett} which is just
adapted namely  for this problem.
For the model (\ref{Ham_Kondo}) this method was described in Ref.\cite{Hewson} where
a Wilson chain for non-discretized  semielliptic density of states was also analytically constructed.
We perform this procedure numerically for an arbitrary logarithmically discretized density of states.
Starting from the vector $|0\rangle = f_{0\sigma}^\dagger |\texttt{vac}\rangle$ (where $f_{0\sigma}^\dagger = \sum_{\eta,m}
\alpha_{\eta m}c_{\eta m,\sigma }^{\dagger}$) we  generate a new basis $|0\rangle, |1\rangle, |2\rangle$... for the
conduction electron states
by Schmidt orthogonalization:
\begin{eqnarray}
|1\rangle &=& \frac{1}{\gamma_0}(H_{\sigma}|0\rangle - |0\rangle\langle 0|H_\sigma|0\rangle) \ , \nonumber \\
|2\rangle &=& \frac{1}{\gamma_1}(H_{\sigma}|1\rangle - |1\rangle\langle 1|H_\sigma|1\rangle - |0\rangle\langle
0|H_\sigma|1\rangle) \ , \\
|n+1\rangle &=& \frac{1}{\gamma_n}(H_{\sigma}|n\rangle - |n\rangle\langle n|H_\sigma|n\rangle - |n-1\rangle\langle
n-1|H_\sigma|n\rangle) \ , \nonumber \label{lanczos}
\end{eqnarray}
where each $\gamma_n$ is chosen to normalize $|n+1\rangle$. One can see that $\langle n'|H_\sigma|n\rangle = 0$ for
$n' = 0, 1, 2... n-2$. This means that $H_\sigma$ is tridiagonal in the new basis. The off-diagonal elements are $\langle
n+1|H_\sigma|n\rangle = \gamma_n$. Defining $\epsilon_n \equiv\langle n|H_\sigma|n\rangle$ we consecutively obtain the
coefficients $\gamma_n$ and $\epsilon_n$ for (\ref{eq:H_chain}).

\subsection*{Calculation of thermodynamical averages}

In our calculations we 
put $\Lambda$ = 1.5 and store $10^4$ states per iteration.

Diagonalizing the Hamiltonian (\ref{eq:H_chain}) for a given chain length $N$ yields a set of eigenvalues. As indicated in
Ref.\cite{Wilson}, because of  retaining only part of the energy spectrum at the $N$-th step of the NRG
procedure, thermodynamic averages should be calculated at a temperature that depends on $\Lambda$: $T_N =
\Lambda^{-N/2}T_0$, where the starting temperature $T_0$ is chosen more or less arbitrarily. However, it should be
neither too large (then the contributions of abandoned high-energy states becomes important), nor too small (then the
discreteness of the energy spectrum becomes appreciable).

The total entropy and  specific heat read \cite{Pruschke}
\begin{eqnarray}
\mathcal{S}_{\rm tot} &=& \langle H\rangle_{\rm tot}/T + \ln Z_{\rm tot} \ , \nonumber \\
C_{\rm tot} &=& \left[\langle H^2\rangle_{\rm tot}-\langle H\rangle_{\rm tot}^2\right]/T^2 \ , \label{SCtot}
\end{eqnarray} where $Z$ is partition function. Since the total spin commutes with the Hamiltonian $H$, each eigenvalue is
characterized by a well-defined spin projection $S_{z,{\rm tot}}$, therefore the quantities like $\langle S_z^2\rangle_{\rm
tot}$ can be calculated straightforwardly. On differentiating $\langle S_z\rangle_{\rm tot}$ with respect to magnetic field
one obtains \cite{Wilson}
\begin{equation}
T\chi_{\rm tot}(T) = \left[ \langle S_z^2\rangle_{\rm tot} - \langle S_z\rangle^2_{\rm tot}\right] \ , \label{Tchitot}
\end{equation} $T\chi_{\rm band}(T)$, $\mathcal{S}_{\rm band}$ and $C_{\rm band}$  are calculated in a similar way, and the
corresponding impurity contributions are obtained by subtracting them from (\ref{SCtot})-(\ref{Tchitot}).

Because of finite length of the chain, the thermodynamic quantities like $T_N\chi_{\rm imp}(T_N)$ and $C_{\rm imp}(T_N)$
demonstrate even-odd oscillations depending on $T_N$ which have nearly constant amplitude. Therefore the amplitude of the
oscillations in  $\chi(T)$ and $C(T)/T=\gamma (T)$ increases strongly with lowering $T$.

To  suppress  the   oscillations in  $\chi(T)$, we used smoothing according to Euler \cite{Hardy}. If there is a certain
oscillating sequence $A_n$, we introduce a new sequence $A^{(1)}_n$ whose members are equal to averages of the adjacent
members of the initial sequence: $A^{(1)}_n= (A_n+A_{n+1})/2$. If necessary, the procedure is repeated: $A^{(2)}_n=
(A^{(1)}_n+A^{(1)}_{n+1})/2$. In particular, this was made in the calculation of $\chi_\mathrm{imp}$. By designating
$\chi_{N} \equiv \chi_\mathrm{imp}(T_N)$, we obtain
\begin{eqnarray}
&\chi^{(1)}(\sqrt{T_NT_{N+1}}) = \frac{1}{2}\chi_{N} + \frac{1}{2}\chi_{N+1}  \nonumber \\
&\chi^{(2)}(T_N) = \frac{1}{4}\chi_{N-1} + \frac{1}{2}\chi_{N} + \frac{1}{4}\chi_{N+1} \\
\nonumber &\chi^{(3)}(\sqrt{T_NT_{N+1}}) = \frac{1}{8}\chi_{N-1} + \frac{3}{8}\chi_{N} + \frac{3}{8}\chi_{N+1}+
\frac{1}{8}\chi_{N+2}
\end{eqnarray}
A similar problem which occurs at calculating the slope of specific heat $\gamma_\mathrm{imp}$ is solved by the same way.
The method was tested for the flat-band case to obtain the values $R=2.008, 2.016$ and $w=0.416, 0.417$ for $J=-0.1, -0.2$
respectively (cf. Table 1). Our method  differs from \cite{Wilson} by that we calculate $\chi$ and $\gamma$ directly rather
than by constructing an effective Hamiltonian explicitly near the fixed point $J=-\infty$. Although resulting in a slight
decrease of accuracy, such an approach can be applied more widely, in particular to obtain a NFL behavior (see
Figs.\ref{Fig_Thi_t0}--\ref{Fig_hiJ02}).



\end{document}